\documentclass[prl, twocolumn, showpacs, floatfix]{revtex4}

\usepackage{graphicx}

\begin{document}

\title{Measurement of the electron electric dipole moment using YbF molecules}

\author{J.J. Hudson}
\author{B.E. Sauer}
\author{M.R. Tarbutt}
\author{E.A. Hinds}
\affiliation{Sussex Centre for Optical and Atomic Physics,
University of Sussex, Brighton BN1 9QH, UK}

\date{\today}

\begin{abstract}
The most sensitive measurements of the electron electric dipole
moment $d_{e}$ have previously been made using heavy atoms. Heavy
polar molecules offer a greater sensitivity to $d_{e}$ because the
interaction energy to be measured is typically 10$^{3}$ times
larger than in a heavy atom. We report the first measurement of
this kind, for which we have used the molecule YbF. Together, the
large interaction energy and the strong tensor polarizability of
the molecule make our experiment essentially free of the
systematic errors that currently limit $d_{e}$ measurements in
atoms. Our first result $d_e = ( - 0.2\pm 3.2)\times 10^{ -
26}\;{\rm e}\,{\rm cm}$ is less sensitive than the best atom
measurement, but is limited only by counting statistics and
demonstrates the power of the method.
\end{abstract}


\maketitle

The permanent electric dipole moment (edm) of an elementary
particle vanishes unless the discrete symmetries parity (P) and
time reversal (T) are both violated \cite{Purcell:1957}. This
naturally makes the edm small in fundamental particles of ordinary
matter. In the standard model of elementary particle physics, a
cancellation of leading order Feynman diagrams further suppresses
the expected value of the electron edm $ d_{e}$ to less than $
10^{ - 38}\;{\rm e}\,{\rm cm}$ \cite{Pospelov:1991}, but in
theories going beyond the standard model, typical predictions are
in the range $10^{ - 26} - 10^{ - 28}\;{\rm e}\,{\rm cm}$
\cite{recent:1999}. The search for an electron edm is therefore a
search for physics beyond the standard model, and particularly it
is a search for non-standard CP violation. CP violation is
intimately connected to T-violation through the CPT theorem
\cite{See:1987}. This is an important and active field at present
\cite{Art:2001} because the prospects for discovering new physics
seem
 so good. Such research is complementary to experiments on K and B particles
where the observed CP violation confirms the standard model CKM
mechanism \cite{Wolfenstein:2001}. The most precise measurement of
the electron edm is that of Regan and Commins \cite{Regan:2002},
who have achieved the result $d_e = (7\pm 8)\times 10^{ -
28}\;{\rm e}\,{\rm cm}$ by searching for a differential Stark
shift between the two hyperfine sublevels $\left( {F = 1,\;m_F =
\pm 1} \right)$ in the ground state of atomic thallium. Their
result imposes stringent constraints on possible theories of
particle physics. Any improvement would make a very important
contribution to the further search for new physics beyond the
standard model, but their method seems to have reached its limit
because of systematic errors associated with stray magnetic
fields. Such errors should be greatly suppressed if $d_e $ could
be measured using heavy polar molecules instead of atoms
\cite{Sandars:1975}, but hitherto this has not been technically
possible. Here we report on the first experiment to realize this
idea, which demonstrates the feasibility of the method and,
although not yet as accurate as the thallium experiment,
nevertheless obtains the next most significant constraint
\cite{unpublished:2002} on $d_e $ . We show that this new
molecular method is potentially much more sensitive than the
previous atomic experiments.

\begin{figure}
\includegraphics[width=3.1in]{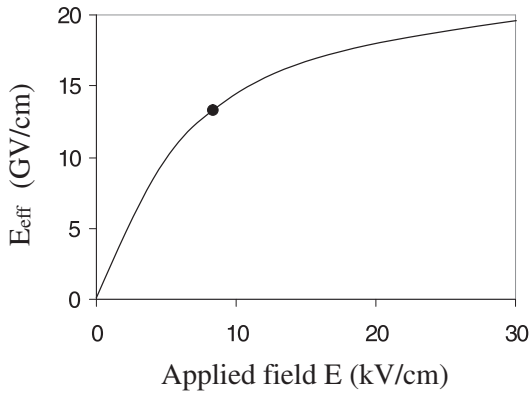}
\caption{Effective electric field interacting with the electron edm in YbF
versus applied electric field. The dot shows the field at which we operate.} \label{fig:field}
\end{figure}

To leading order in $d_e $ the edm interaction of an electron in an atom or
molecule can be described by an effective Hamiltonian $d_e (1 - \gamma _0
)\Sigma \cdot {\rm {\bf E}}_{tot} $, where $\gamma _0 $ and $\Sigma $ are
standard Dirac matrices and ${\rm {\bf E}}_{tot} $ is the total electric
field at the position of the electron, including the field of the other
electrons \cite{See:1}. The explicit form of this interaction,
\begin{equation}
V = \left\langle {\psi _0 \left| {\begin{array}{*{20}c}
   {0} & 0  \\
   {0} & {2d_e \hat{\sigma}  \cdot {\bf E}_{tot} }  \\
\end{array}} \right|\psi _0 } \right\rangle
\end{equation}
where $\hat {\sigma }$ is a unit vector along the electron spin,
shows that the effect is relativistic: the operator connects only
the small components of the Dirac wavefunction $\left| {\psi _0 }
\right\rangle $. The sensitivity to $d_e $ is therefore largest in
heavy systems where the electrons are most relativistic. If there
is no applied electric field, the small component of $\left| {\psi
_0 } \right\rangle $ has definite parity and therefore the
expectation value of this odd-parity operator vanishes. However,
an applied field $E\,{\rm {\hat {z}}}$ polarizes the wavefunction
and yields an energy $V$ of the form $ - d_e E_{eff} \left\langle
{\hat {\sigma } \cdot {\rm {\hat {z}}}} \right\rangle $. A strong
laboratory field applied to an atom induces only a small mixing of
atomic orbitals and therefore $E_{eff}$ is essentially
proportional to $E$, with the ratio scaling roughly as $8Z^3\alpha
^2$ \cite{Hinds:1997}, $Z$ being the nuclear charge and $\alpha $
the fine structure constant. The cesium atom used by Hunter's
group to measure $d_e $ \cite{Murthy:1989} is a particularly
favorable case with the applied field being effectively amplified
by 119 \cite{recent:1999}. Thallium has an even larger
amplification of --585, a key factor in the more sensitive result
of ref. \cite{Regan:2002}. Diatomic polar molecules have a great
advantage over atoms \cite{Sandars:1975} because the atomic
orbitals are naturally strongly polarized along the internuclear
axis $\hat {\lambda }$. This gives rise to an internal effective
electric field $E_{int} \hat {\lambda }$ that can be very large
when one of the atoms is heavy and has strong s-p hybridization of
its orbitals. The molecular rotation averages this strong field to
zero unless an external field $E\,{\rm {\hat {z}}}$ is applied.
However, only a modest field is needed to polarize $\hat {\lambda
}$ along
 $\,{\rm {\hat {z}}}$ because this involves mixing rotational
states, which are much closer together than the electronic states
of an atom. The edm interaction energy $V$ once again takes the
form $ - d_e E_{eff} \left\langle {\hat {\sigma } \cdot {\rm {\hat
{z}}}} \right\rangle $, but $E_{eff} $ is now given by $E_{int}
\left\langle {\hat {\lambda } \cdot {\rm {\hat {z}}}}
\right\rangle $. Figure~\ref{fig:field} shows $E_{eff} $ of the
YbF molecule, rising towards its asymptotic value $E_{int} $ of
26~GV/cm \cite{Kozlov:1998} as the applied field is increased. In
our experiment the applied field is 8.3~kV/cm, for which $E_{eff}
$ is 13~GV/cm. Such a large effective field is not particular to
YbF but can be found in a variety of other heavy polar diatomic
molecules, some of which are listed in Table~\ref{tab:effFields}.
In short, the edm interaction in heavy polar molecules can be
thousands of times larger than in heavy atoms.

\begin{table} [b]
\begin{tabular}{c p{20pt} c}
\hline \hline Species: state&&
$E_{eff}$ (GV/cm) \\
\hline BaF: X$^{2}\Sigma ^{ + }$&&
7.4$^{(a)}$ \\
YbF: X$^{2}\Sigma ^{ + }$&&
26$^{(b)}$ \\
HgF: X$^{2}\Sigma ^{ + }$&&
99$^{(c)}$ \\
PbF: X$^{2}\Sigma ^{ + }$&&
-29$^{(c)}$ \\
PbO: a(1)$^{3}\Sigma ^{ + }$&&
6$^{(d)}$ \\
\hline \hline
\end{tabular}
\caption{Effective electric fields for some heavy polar molecules.
$^{a}$ref. \cite{Kozlov:1997}, $^{b}$ref. \cite{Kozlov:1998},
$^{c}$ref. \cite{Dmitriev:1992}, $^{d}$ref. \cite{DeMille:2000}}
\label{tab:effFields}
\label{tab2}
\end{table}

\begin{figure} [t]
\includegraphics[width=3.1in]{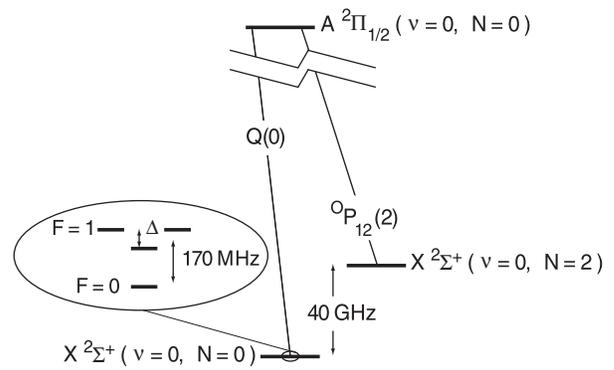}
\caption{Important optical transitions Q(0) and $^{\rm
O}$P$_{12}$(2) in $^{174}$YbF at 553~nm. They are 40~GHz apart.
Inset: ground state hyperfine levels $F = 0$, $F = 1$, 170~MHz
apart. In static electric field, the $m_F = 0$ sublevel of $F = 1$
is lower than the $m_F = \pm 1$ sublevels by an amount $\Delta $.}
\label{levels}
\end{figure}

Our experiment uses $^{174}$YbF in the electronic, vibrational and
rotational ground state $X^2\Sigma ^ + (v = 0,\;N = 0)$. The
electron spin (1/2) and the fluorine nuclear spin (1/2) combine to
produce a hyperfine singlet $F = 0$ and triplet $F = 1$, separated
by 170~MHz \cite{Sauer:1995} as shown inset in fig.~2. The applied
electric field lowers the energy of the $\left| {F,m_F }
\right\rangle = \left| {1,0} \right\rangle $ state relative to
$\left| {1,\pm 1} \right\rangle $ by an amount $\Delta $ ($\Delta
/ h = 6.7$~MHz for the 8.3~kV/cm used in our experiment)
\cite{Sauer:1996}. This strong tensor splitting reflects the
non-spherical symmetry of the molecule's internal structure. The
two states $\left| {1, + 1} \right\rangle $ and $\left| {1, - 1}
\right\rangle $ remain degenerate as a consequence of
time-reversal symmetry. Their degeneracy is lifted by the edm
interaction, which causes a splitting of $2d_e E_{eff} $ that we
seek to measure in the experiment. A magnetic field small compared
with $\Delta / \mu _B $ causes an additional splitting
\cite{Hudson:2001} of $2 \mu _B B_z (1-\frac{1}{2}(\mu _B B_ \bot
/ \Delta)^2)$ plus higher order corrections (here the g-factor,
both expected and measured, is 1). This formula shows that the
field parallel to \textbf{E} induces a Zeeman splitting $2\mu _B
B_z $ between the $m_F = \pm 1$ sublevels, whereas the splitting
due to the perpendicular field $B_ \bot $ is suppressed relative
to $\mu _B B_ \bot $ by a factor $\mu _B^2 B_z B_ \bot / \Delta
^2$, which is $3\times 10^{ - 10}$ in our experiment ($B_z \approx
10\;{\rm nT}$, $B_ \bot \approx 6\;{\rm nT})$. We separate the
splitting due to the edm interaction from that of the magnetic
interaction by reversing the directions of the applied electric
and magnetic fields, \textbf{E }and \textbf{B}. The edm part of
interest has the symmetry of ${\rm {\bf E}} \cdot {\rm {\bf B}}$,
as one might expect for a P-odd, T-odd effect. The suppression of
the splitting induced by $B_ \bot $ is a critical aspect of the
experiment because the motion of the molecules through the
electric field generates a 6~nT contribution to $B_ \bot $,
$B_y^{mot} = Ev / c^2$, which reverses with \textbf{E} and
therefore has the potential to masquerade as an edm
\cite{Sometimes:1995}. If $B_ \bot $ is entirely motional it does
not generate a false edm because the splitting depends on $B_ \bot
^2 $, remaining unchanged when $B_ \bot $ reverses. However, if
there is also a small $y$-component $B_y $ of the applied magnetic
field, the magnitude of $B_ \bot $ will change when either
\textbf{E} or \textbf{B} is reversed, leading to an apparent edm
given by $\mu _B^3 B_z B_y^{mot} B_y / \Delta ^2E_{eff} $. In our
experiment $B_y $ is less than 1~nT and therefore this false
$d_{e}$ is less than $ 10^{ - 33}\;{\rm e}\,{\rm cm}$. The
advantage of a strong tensor polarizability for edm measurements
was first demonstrated by Player and Sandars using the $^3P_2$
metastable state of Xe \cite{Player:1970}. These two features of
heavy polar molecules
--- large $E_{eff} $ and strong tensor polarizability
--- give them such excellent suppression of all the known systematic errors
that a major improvement in $d_{e}$ now seems accessible.

\begin{figure}
\includegraphics[width=3.2in]{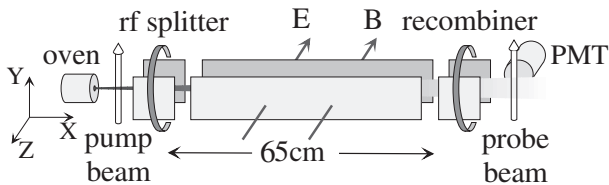}
\caption{Schematic of the YbF molecular beam apparatus.}
\label{schematic}
\end{figure}

Our YbF molecular beam, illustrated in fig.~3, effuses out of a
molybdenum oven containing a mixture of Yb metal and powdered
AlF$_{3}$ (mass ratio 4:1) heated to $\sim $1500K. The molecules
are detected by dye-laser-induced fluorescence 1~m away from the
source, using collection optics with 22{\%} efficiency and a
photomultiplier (PMT) of 10{\%} quantum efficiency. The detection
laser is tuned to the $F = 0$ component in the Q(0) line of the
A$^{2}\Pi _{1 / 2}$-X$^{2}\Sigma ^ + $ electronic transition
(fig.~2) at 553~nm. This induces a count rate of 2.5~kHz due to
the $^{174}$YbF ground state $F = 0$ molecules and 20~kHz from
nearby transitions involving higher rotational states and other
isotopes of Yb \cite{Sauer:1996}. There are also backgrounds of
$\sim $60~kHz from oven light and $\sim $30~kHz from scattered
laser light. The pump beam of fig.~3 consists of two dye laser
beams.  One is a 170~MHz red sideband of the probe laser. This
removes molecules from all the sublevels of the  $F = 1$ ground
state by exciting them to A$^{2}\Pi _{1 / 2}$, which has
negligible hyperfine structure \cite{Sauer:1999}. A second dye
laser 40~GHz to the red, tuned to the $^{\rm O}$P$_{12}$(2)
transition (fig.~2), excites molecules out of the highest
hyperfine level of the $N = 2$ rotational manifold.  Together
these two beams pump molecules into the $F = 0$ ground state,
increasing the probe signal to 8~kHz, while emptying out $F = 1$.
After pumping, the beam enters an electric field of 3.3~kV/cm,
where a 170~MHz rf magnetic field along the $x$-direction excites
the molecules to a coherent superposition $\textstyle{1 \over
{\sqrt 2 }}\left| {1, + 1} \right\rangle + \textstyle{1 \over
{\sqrt 2 }}\left| {1, - 1} \right\rangle $ of the $\left| {F,\;m_F
} \right\rangle $ states. In essence this step is the beam
splitter of a molecular interferometer. Next the molecules enter
the center of the interferometer, where they evolve for a time $T$
($\sim $1~ms) in combined electric and magnetic fields $(\pm E,
\pm B){\, \rm { \hat {z}}}$, and the two parts of the wavefunction
develop a relative phase shift of $2\varphi = 2\left( {\pm d_e
E_{eff} \mp \mu _B B} \right)T / \hbar $. To measure this phase, a
second rf loop drives a transition back to the $F = 0$ state. This
is the recombiner in fig.~3.  The state amplitude $\textstyle{1
\over {\sqrt 2 }}(e^{i\varphi } + e^{ - i\varphi })$ gives a final
$F = 0$ population proportional to $\cos ^2\varphi $. Our
detector, measuring fluorescence from molecules excited on the
Q(0),~$F = 0$ transition, probes this population. Fig.~4 shows the
central interferometer fringe, measured by scanning the applied
magnetic field. To obtain this fringe pattern it is necessary to
suppress stray magnetic fields, which we do by means of two
1~mm-thick layers of high-permeability (Ad-MU 80) magnetic
shielding. We note that coherences at the 170~MHz rf frequency
play no role in this experiment; indeed the two rf loops are
driven by 2 different oscillators to avoid any such Ramsey
interferences.

\begin{figure}
\includegraphics[width=3.1in]{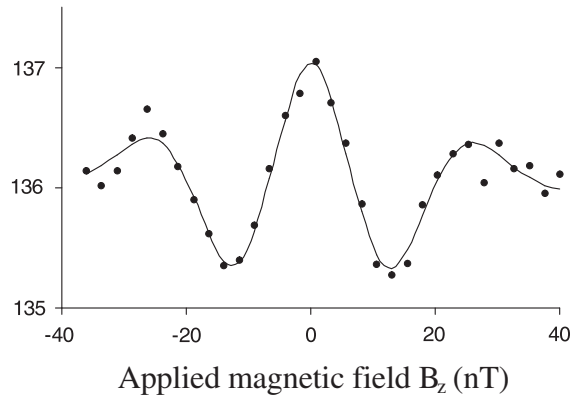}
\caption{Interference fringe in the $F = 0$ state population versus magnetic
field. Each point represents 12~s of integration time. The curve is a
velocity-averaged calculation whose only free parameters are the
normalization, a field offset, and a sloping background.} \label{fringes}
\end{figure}

The interferometer is most sensitive to small changes in phase
when $\varphi = \pm \pi / 4$, i.e. when the magnetic field $\pm
B_0 $ is set near the points of steepest slope at $\pm $6~nT on
either side of the central fringe. We record the fluorescence in
the detector under the 8 conditions $\pm E,\;\pm (B_0 \pm \delta
B)$, where $\delta B$ is a carefully calibrated change of 1.6~nT
in the magnitude of the applied field. Over a period of 82 seconds
these 8 measurements are repeated 128 times, ordered in such a way
as to minimise drifts \cite{Harrison:1971}. The duty factor is
60{\%}. This constitutes a ``block'' of data. In order to maintain
operation near the most sensitive points (the residual field
drifts by a few~nT over a period of hours), we compare the PMT
count rates for $ + B_0 $ and $ - B_0 $ (averaged over the other
reversals) at the end of each block. Any difference indicates a
nonzero residual field, which we null by adding a small
proportional bias.

The dye in both lasers has a short lifetime ($\sim $20 hours) as
does the molecular beam source ($\sim $10 hours). Frequent
maintenance and the small signal and large background in our
detector make the experiment difficult to operate at present.
Nevertheless, we have succeeded in taking 24 hours of data over a
period of 4 months, a total of 1758 data blocks containing
$1.1\times 10^{10}$ counts. In our analysis of the data we extract
the change in PMT count rate correlated with every combination of
$E$, $B$ and $\delta B$ reversals within every block. Two of these
combinations give the phase shift due to the edm interaction and
the phase shift due to $\delta B$. Their ratio, $d_e E_{eff} / \mu
_B \delta B$, yields the edm result $d_e = \left( { - 0.2\pm 3.2}
\right)\times 10^{ - 26}\;{\rm e}\,{\rm cm}$, where the
uncertainty (1$\sigma$) is due entirely to random pulse counting
statistics.

All other combinations of all the switched parameters are
consistent with zero, suggesting that systematic errors are well
under control. As a further check, we took the data in 4 sets with
the connections to the electric plates and/or magnetic field wires
reversed, and this too revealed no systematic errors. We monitored
the leakage current of the plates, which was always below 10~nA on
both plates during data taking. At worst, this current
(implausibly looping around the edge of each plate) would cause a
false edm of $4\times 10^{ - 28}\;{\rm e}\,{\rm cm}$. This is
small compared with the result of ref \cite{Regan:2002} and is
insignificant in the present experiment. With a view to future
improvements in sensitivity, we checked the effect of a change in
field magnitude when the electric field was reversed. This changed
the resonance strength and therefore the slope of the fringe. When
combined with an asymmetry of the magnetic field it led to a false
edm, but one that we were fully able to understand and correct on
the basis of the separate E-reversal and B-reversal signals. We
also ran the experiment with $ B_ \bot $ intentionally amplified
by a factor of 50 and saw that this did not induce any false edm
at the level of $5\times 10^{ - 26}\;{\rm e}\,{\rm cm}$. All these
effects were negligible in the present experiment and seem to be
well under control for a future measurement below $1\times 10^{ -
27}\;{\rm e}\,{\rm cm}$. Some details of the systematic checks are
available in reference \cite{Hudson:2001}.

This experiment has demonstrated that a molecular measurement of
$d_e $ is less sensitive to stray magnetic fields than any
previous experiment by several orders of magnitude. This is due to
the high value of $E_{eff} $. The cylindrical symmetry of the
electronic state also makes the molecule immune to the motional
magnetic field effect. The present result could improve
considerably with higher beam intensity. To this end, we have
developed a new pulsed source in which YbF molecules are produced
by laser ablation and cooled below 20~K by supersonic expansion in
a jet of carrier gas.  This produces 100 times more population in
the $N = 0$ rotational state. On the basis of its current
performance, we expect that incorporating the supersonic source
into our experiment will reduce the statistical uncertainty in
$d_{e}$ to below $1\times 10^{ - 27}\;{\rm e}\,{\rm cm}$ in 24
hours. A competing experiment is in preparation at Yale University
by the group of D. DeMille using a cell of PbO vapor excited to
the paramagnetic metastable a(1) state \cite{DeMille:2000}.
Trapped cold atoms offer a possible alternative to using polar
molecules for measuring $d_e $ \cite{Chin:2001}. They can have
transverse spin relaxation times of seconds instead of
milliseconds and the atoms move slowly making the ${\rm {\bf
E}}\times {\rm {\bf v}}$ effect much less than in an atomic beam.
Even so, the sensitivity to stray magnetic fields and to the
trapping fields is a very severe problem for paramagnetic atoms
and the experiment would have to use two trapped species, a heavy
one for measuring $d_e $ and one for monitoring the magnetic
field. Trapped cold molecules offer the best prospect of all, but
at present the required techniques are only in a very early stage
of development \cite{Levi:2000}.

We are indebted to Alan Butler for his expert technical
assistance. This work was supported by the EPSRC and PPARC
research councils of the UK.

\end{document}